# Raman Studies on $MgB_{2-x}C_x$

**T. Sakuntala#, S.K. Deb#, A. Bharathi, S.Jemima Balaselvi, C. S. Sundar and Y.Hariharan**
#Solid State Physics Division, Bhaba Atomic Research Centre, Mumbai, 400086
Materials Science Division, Indira Gandhi Centre for Atomic Research, Kalpakkam, 603102

***Abstract***

Raman scattering measurements have been carried out in $MgB_{2-x}C_x$ for x=0.0 - 0.30. The broad Raman band around 600 $cm^{-1}$ range which correspond to the $E_{2g}$ mode in $MgB_2$, is seen to narrow and reduce in intensity with increasing carbon content. For x > 0.15, new set of bands appear at 1200$cm^{-1}$ possibly arising from the B-C, $E_{2g}$ modes. The decrease in $T_C$ with carbon substitution seems to correlate with this change over.

## INTRODUCTION

Following the discovery [1] of superconductivity at 39 K in $MgB_2$, extensive investigations on the normal and superconducting properties, including several studies on doped systems have been carried out (for a review see [2]). It is now widely believed that superconductivity in the $MgB_2$ system arises due to the coupling of holes in the covalent σ band with the $E_{2g}$ optical phonon, involving the in-plane B-B bond stretching vibrations [3]. A definitive indication of the strong electron-phonon coupling of the $E_{2g}$ mode with the carriers is provided by Raman scattering experiments, as exemplified by the studies [4] on $Mg_{1-x}Al_xB_2$.

Our earlier investigations [5] on the carbon doped system, $MgB_{2-x}C_x$ showed that $T_c$ decreases systematically from 39 K to 26 K as the carbon content increases for x=0 to x=0.5. An interesting aspect of the reduction of $T_c$ with x is a distinct increase in slope at x = 0.15. Qualitatively, the decrease in $T_c$ can arise due to reduction in the hole density due to doping, and /or changes in the phonon spectrum brought about by carbon doping. Analysis of the temperature dependence of resistivity in these samples, in terms of Bloch-Gruneisen formalism, points to a reduction in the Debye temperature [5], that can account for the decrease in $T_c$. Recent theoretical calculations [6] indicate a monotonic reduction in the density of states at the Fermi level with carbon content, that leads to a progressive reduction in $T_c$, with superconductivity ultimately vanishing for x = 0.6. Given these various possibilities for the reduction in $T_c$, it is of interest to know if there are any changes in the $E_{2g}$ mode with carbon doping, and it is with this motive the present Raman investigations have been undertaken.

## EXPERIMENTAL

The samples of $MgB_{2-x}C_x$ were prepared by the solid-vapour route under 50 bar locked-in Ar pressure as described in Ref. [5]. Raman spectra were recorded from as-prepared chunks as well as pellets. The 532 nm laser line, of power 15 mW was used to excite the Raman line. Scattered light was detected using a CCD based, home built Raman spectrometer together with a super notch filter, covering the range of 200 –1750 $cm^{-1}$. On each sample 10-12 measurements were made on different spots.

## RESULTS AND DISCUSSIONS

Fig. 1 shows the Raman spectra of $MgB_{2-x}C_x$ samples for some representative concentrations x. Spectrum of the pure $MgB_2$ could be fitted to a sum of two Gaussians centered at 560 and 740 $cm^{-1}$ of widths 250 and 300 $cm^{-1}$ respectively. The former can be identified with the $E_{2g}$ phonon [4] and the latter is understood to be arising due to the sampling of the phonon density of states due to disorder [7]. With increasing carbon concentration, it is noted that the intensity of the band around 740 $cm^{-1}$ picks up. For compositions beyond x = 0.15, qualitative differences in the Raman spectra are seen. In particular, it is noted that as the mode at 560 $cm^{-1}$ disappears and a broad feature around 1200 $cm^{-1}$ develops, that starts gaining in strength. Further, the Raman signal for samples with x ≥ 0.15 is also found to be much larger.

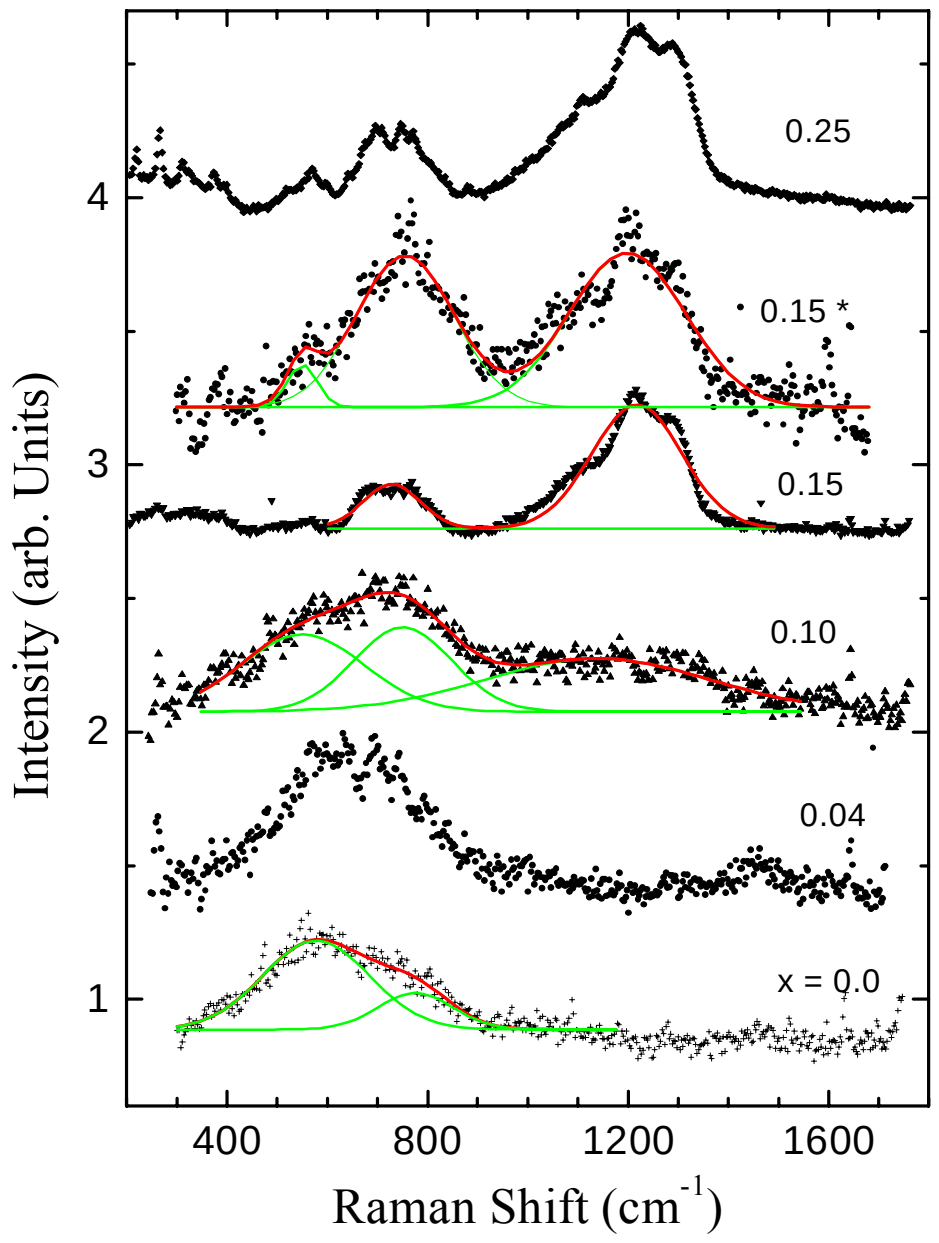

**Fig. 1**. Raman Spectra of $MgC_xB_{2-x}$ for different x. Solid lines shows the fit to the data as a sum of gaussians. For x = 0.15, the spectra correspond to different spots on the sample.

A careful examination of these Raman spectra reveal that for samples with x ≥ 0.15, the line shape resembles, rather closely, the phonon density of states. This reflects the build up of disorder, which in the present case is due to the incorporation of carbon in the boron plane. The feature at 1200 $cm^{-1}$ can be associated with the B-C bond stretching vibrations. This identification can be made based on recent studies on an isostructural system, $Li_xBC$, that shows the $E_{2g}$ Raman mode at 1176 $cm^{-1}$ and sharp features in the phonon density of states in this spectral region [8,9]. The hardening of the B-C vibrations as compared to the B-B vibrations is one of the reasons for expecting higher superconducting transition temperature [8] in the $Li_xBC$ system.

Fig.2 shows the evolution of the position and width of the Raman bands as a function of carbon fraction. The top panel shows that with increasing x, upto x =0.1, the frequency of the $E_{2g}$ phonon is found to marginally increase and this is associated with a reduction in the phonon linewidth. While the increase in the mode frequency can be attributed to the observed reduction in the 'a' lattice parameter [5], the observed mode hardening, associated with a reduction in the linewidth points to a reduction in the electron-phonon coupling of the $E_{2g}$ mode. Allen's formula [10] relates the phonon linewidth $\Gamma_i$ to the electron-phonon coupling constant $\lambda_i$ through

$$\Gamma_i = (2\pi/g_i)\, \lambda_i\, N(0)\, \omega_i^{\,2}$$

where, $g_i$ $\omega_i$ and N(0) are the degeneracy of the mode, the bare- phonon frequency the density of states at the Fermi level respectively. From the observed variation in linewidths of the $E_{2g}$ mode and using the value of 980 $cm^{-1}$, for the bare phonon frequency [4], the variation in $\lambda_i$N((0) with carbon doping has been estimated using Allen's formula. This indicates a reduction in $\lambda_i$N((0) by ~ 30 % as the carbon content increases from x=0 to x=0.10. Electronic structure calculations by Singh [6], indicate a linear decrease in the in the density of states from 10.5 states / Ryd/ atom for x=0 to 7.5 states / Ryd/ atom for x= 0.4, and this translates to a reduction in N(0) by ~ 7 %, as the carbon content increases from x=0 to x=0.1. Thus, we note that the calculated decrease in $\lambda_i$N((0) ( ~ 30%) cannot be solely accounted for in terms of a decrease in N(0) alone, implying that there is also an appreciable decrease in the electron - phonon coupling strength, $\lambda_i$ with carbon doping. It is well known [11] that the interaction of phonon with the electronic continuum, apart from resulting in an increase in the phonon line width also leads to a strong decrease in the phonon frequency, when compared to the bare phonon frequency. The extent of down shifting scales [11] with the coupling strength , $\Delta\omega_i/\omega_i \sim -(\lambda_i / g_i)$ . Thus, the hardening of the $E_{2g}$ mode frequency by ~ 10 % seen in Fig.2 also points to and is consistent with the reduction in the coupling strength $\lambda_i$ with carbon doping.

It is seen from Fig. 2 that apart from the changes in the $E_{2g}$ mode, the features corresponding to the phonon density of states also show significant changes with carbon doping. In particular, these features seem to harden initially with a corresponding decrease in width. This may be due to the shrinkage in the "a" lattice parameter [5] with C doping.

Our earlier superconductivity studies $MgB_{2-x}C_x$ had shown [5] that Tc initially drops slowly from a value of 40 K for x=0 to 35 K for x=0.15, and then sharply to 25 K for x=0.4. A significant result from the present study is that there are qualitative differences in the Raman spectra (see Fig.1) at the same composition (x = 0.15) at which the slope of the $T_c$ vs x is observed to increase [5]. It is also interesting to note that carbon doped samples, for $x \geq 0.15$, is superconducting at all with a $T_c$ as high as 25 K, though the $E_{2g}$ mode has manifestly disappeared, as seen in Figs 1 and 2. We surmise that superconductivity for these compositions is sustained mainly by the B-C vibrations, centred around 1200 $cm^{-1}$. The present results may have bearing on the attempts [8,9] to make Li deficient, $Li_xBC$ system superconducting. It suggests that that $Li_xBC$ has not been made suprconducting so far, not because the B-C vibrations do not couple significantly with the hole-carriers, but that adequate holes are not getting doped into the B-C layer.

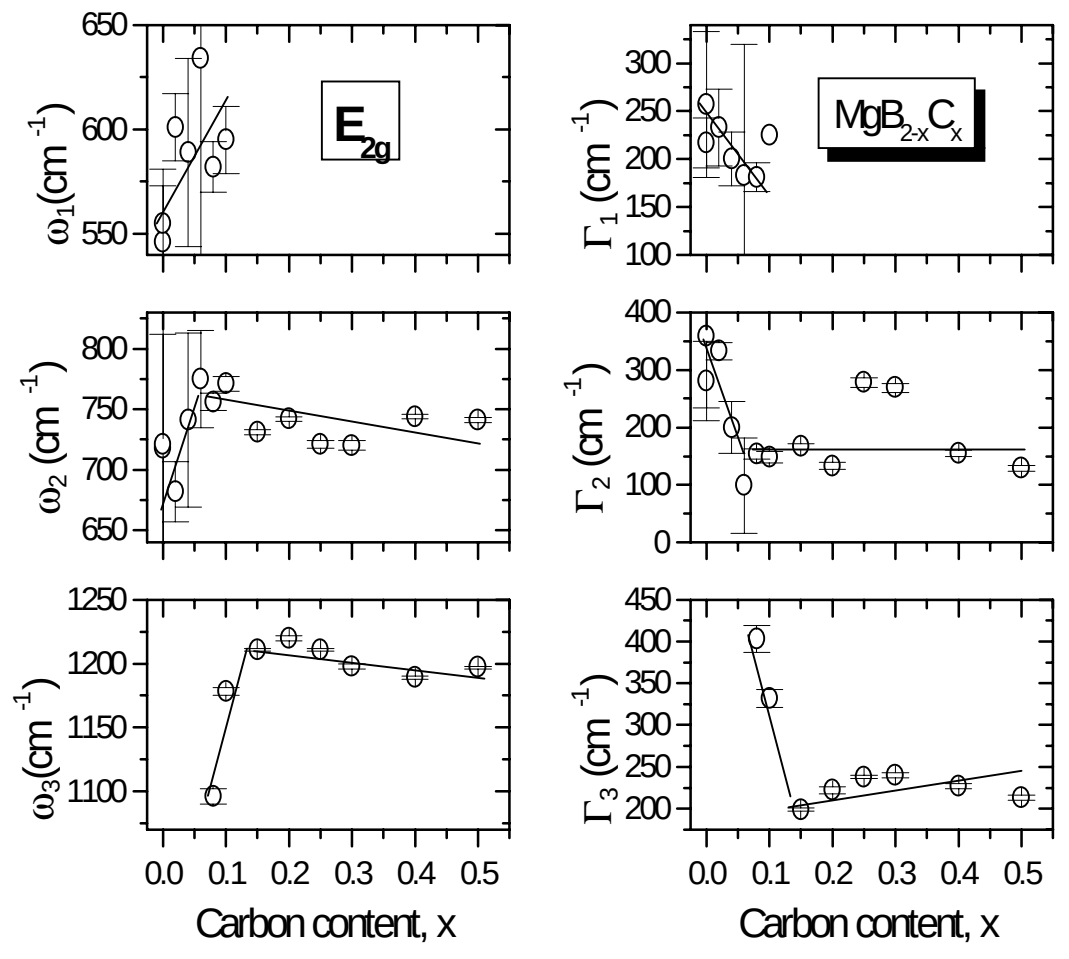

**Fig.2**. Composition dependence of the frequency and line width of Raman bands